\documentclass[11pt,twoside,a4paper]{article}
\usepackage[affil-it]{authblk}
\usepackage{amssymb}
\usepackage{amsmath}
\usepackage{tensor}
\usepackage{color}
\usepackage{tikz}
\usetikzlibrary{matrix}
\usepackage[margin=2.9 cm]{geometry}
\usepackage{polski}
\usepackage[cp1250]{inputenc}
\usepackage[polish,english]{babel}
\usepackage[toc,page]{appendix}
\usepackage{amsthm}
\usepackage{stmaryrd}

\usepackage{hyperref}
\input xy
\xyoption{all} \tolerance=500

\def\sT{\mathsf T}
\newdir{ (}{{}*!/-5pt/@^{(}}
\newdir{|>}{!/4,5pt/@{|}*:(1,-.2)@^{>}*:(1,+.2)@_{>}}

\def\mg{\mathfrak g}
\def\mh{\mathfrak h}
\def\mj{\mathfrak j}
\def\d{\mathrm d}

\def\sV{\mathsf V}

\def\sJ{\mathsf J^1}

\def\J2{\mathsf J^2}
\def\sj{\mathsf j^1}
\def\ind{\indices}

\def\Ad{\mathrm{Ad}}
\def\ad{\mathrm{ad}}

\def\ddt{\frac{\mathsf d}{\mathsf dt}_{|t=0}}

\def\cV{\mathcal V}
\def\cF{\mathcal F}
\def\C2{\mathbb C^2}
\def\barP{\overline{\mathcal P}}

\def\wdt{\widetilde }

\title{The dressing field method in gauge theories - geometric approach}

\begin{document}

\author{Marcin Zaj\k{a}c}
\date{}
\maketitle
\vskip -1.0cm

\centerline{Department of Mathematical Methods in Physics,}
\centerline{Faculty of Physics. University of Warsaw,}
\centerline{ul. Pasteura 5, 02-093 Warsaw, Poland.}
\centerline{marcin.zajac@fuw.edu.pl}

\vskip 1.0cm

\begin{abstract}
Recently, T. Masson, J. Francois, S. Lazzarini, C. Fournel and J. Attard have introduced a new method of the reduction of gauge symmetry called a dressing field method. In this paper we analyse this method from the fiber bundle point of view and we show the geometric implications for a principal bundle underlying a given gauge theory. We show how the existence of the dressing field satisfying certain conditions naturally leads to the reduction of the principal bundle and, as a consequence, to the reduction of the configuration and phase bundle of the system.

\end{abstract}

\tableofcontents

\section{Introduction}



\qquad Gauge field theories form a theoretical basis of the modern understanding of fundamental interactions and the research on them is in the very center of theoretical physics. Perhaps the most prominent example of gauge theories are Yang-Mills theories, which provide a description of electroweak and strong interaction between elemetary particles.
From the mathematical point of view, gauge field theory found its mathematical formulation in the geometry of principal bundles. Usually, one takes a principal bundle $\pi:P\to M$ with a structural group $G$, where $M$ represents a physical spacetime. The crucial object here is a connection in a principal bundle, which may be defined by means of a one form $\omega: \sT P\to\mg$, where $\mg$ is a Lie algebra of the group $G$. From the physical point of view connections in $P$ represent gauge fields of the given theory. Therefore, the space of gauge fields is the space of connections in a principal bundle $\pi:P\to M$. On the other hand, the Lagrangian of the system in classical field theory is usually a map $\sJ E\to\Omega^m$, where $\tau:E\to M$ is a bundle of fields, $\sJ E$ is a first jet bundle of $\tau$ and is a space of $m$-covectors on $M$. Here, we assumed that the dimension of the base manifold $M$ is $m$. Therefore, one has to consider first jets of the connection form when analysing the structure of gauge theory. These two structures, namely principal bundles and jet bundles, form a geometrical basis of the description of gauge field theories. In our paper we will not focus on the derivation of the dynamics but rather on the description of kinematics, i.e. on the fibre bundles appearing in gauge theory.

However, despite its numerous successes, mathematical description of the gauge fields still encounters certain significant problems. Perhaps the most serious one is the fact that the existence of a gauge symmetry makes the quantization of such a theory very problematic. Therefore, one of the main problems of gauge theories is the problem of finding a proper way to reduce the gauge symmetry of the system.

There are three main aproaches to the reduction of the gauge symmetry existing in the literature. The most popular in experimental physics and the simplest one from the conceptual point of view is the so-called gauge fixing. Since the system has a gauge symmetry, all the fields belonging to the same orbit of the gauge transformation describe the same physical state. It means that we can "by hand" choose a particular gauge in such a way that e.g. the calculations take the simplest possible form. The second approach is based on a mechanism of the spontenous symmetry breaking. This approach is a basis of the famous Higgs mechanism, which solved the problem of the mass generation for bosons carrying electroweak interaction. The last one, and the most geometric one, is related to the bundle reduction theorem. It turns out, that in certain situations, the principal bundle with a given structural group may be reduced to its subbundle with a smaller structural group.



In our paper we will focus on a new method of reduction of a gauge symmetry which is called a dressing field method. This approach has been recently discovered mainly by J. Attard, J. François, S. Lazzarini and T. Masson \cite{AFM,FFLM,MW,FLM}. The main idea of this approach is to introduce into a theory certain auxiliary field, which does not belong to the original space of fields of the given gauge theory. In the next step one performs a change of variables in which the original gauge fields are transformed into new fields, which are combination of the original gauge fields and the dressing field. This procedure is known as a dressing of the gauge field. The new fields, called dressed fields, are the new variables of the theory. In favorable situations these new fields are invariant under the action of the subgroup of the structural group $G$. It means that the symmetry related to this subgroup has been erased.
Let us stress that the authors approach to the entire method is rather algebraic and they do not delve into the geometric character of the dressing procedure. Our attitude is, in a sharp contrast, strictly geometric with the principal bundle and jet bundle geometry as a basis.



The aim of our paper is to explore the dressing field method from the geometrical point of view. Since the basis of the mathematical formulation of gauge theories are principle bundle geometry and jet spaces, we show how the existence of the dressing field affects both geometries. 
The important tool in our work will be the decomposition of the jet bundle $\sJ C$ that may be found in \cite{SG2}. In our work we proceed in three steps. First we assume that the structural group of the theory has a distinguished subgroup $H$ together with a dressing field $u:P\to H$. In the second step we assume that there is a decomposition $G=JH$, where $H\subset G$ is a normal subgroup of $G$. In the last step we analyse the case $G=H\times J$, i.e. when the structural group is a direct product of its subgroups. The last step is particularly important in the reduction of the gauge symmetry in the electroweak theory, where the underlying group is $G=SU(2)\times U(1)$.

The paper is organised as follows. In section 2 we review the geometric fundamentals of gauge field theories. We recall the basics of principal bundles and the notion of a connection and its curvature in a principal bundle. In particular we show how the connection in $\pi:P\to M$ may be represented as a suitable section of the first order jet bundle $\sJ P\to P$. Section 3 contains a coincise presentation of the concept of gauge transformations with three different pictures of this notion. In section 4 we briefly introduce the reader to the dressing field method basing mainly on \cite{AFM}. Section 5 is the core of our paper. 
Here we discuss the geometric interpretation of the existence of the dressing field on $P$. We show that, under suitable assumptions, the existence of the dressing field implies that the original principal bundle may be reduced to a smaller principal bundle with a structural group isomorphic to $G/H$. This reduction naturally leads to the simplification of the entire structure of the given gauge theory. It turns out that the level of the possible reduction strongly depends on the structure of the structural group $G$, the issue which is discussed by us in detail. 


\section{Geometric formulation of gauge theories}

In this section we will briefly recall the geometry of principal bundles and the first order jet bundles. We will also fix the notation necessary for our subsequent work. The main element of this chapter is the subsection 2.4 where we show how to represent the connection in the principal bundle as a section of the proper jet bundle. The introduction concerning principal bundles and connections in them is in large based on \cite{JMFF}.

\subsection{Principal bundles and adjoint bundles}

Let $G$ be a Lie group with a Lie algebra $\mg$. We denote by $P$ a smooth manifold such that $G$ acts on it from the right-hand side in a smooth, free and proper way. Then, the space $M:=P/G$ of orbits of the action of $G$ on $P$ is a smooth manifold as well. The bundle $\pi:P\to M$ is called a principal bundle with a structural group $G$. It is locally isomorphic to $M\times G$. Let $U_\alpha\subset M$ be an open subset in $M$. A local trivialisation of a principal bundle is a $G$-equivariant diffeomorphism 
$$  \Psi_\alpha:\pi^{-1}(U_\alpha)\to U_\alpha\times G, \qquad   \Psi_\alpha(p)=(\pi(p),g_\alpha(p)),$$ 
where $g_\alpha:\pi^{-1}(U_\alpha)\to G$ is a $G$-valued function associated with the map $\Psi_\alpha$. The equivariance condition means that $\Psi_\alpha(pg)=\Psi_\alpha(p)g$, which implies that $g_\alpha$ is also $G$-equivariant, says $g_\alpha(pg)=g_\alpha(p)g$. Notice, that the function $g_\alpha$ uniquely defines a local trivialisation of $P$. The transition between trivialisations $g_\alpha$ and $g_\beta$ defined on $\pi^{-1}(U_\alpha\cap U_\beta)$ is realised by the function  
$$  g_{\alpha\beta}:U_\alpha\cap U_\beta\to G,   \qquad  g_{\alpha\beta}(\pi(p))=g_\alpha(p)g_\beta(p)^{-1}.   $$
Let $F$ be a smooth manifold and let $G$ act on $F$ from the left-hand side. We introduce the action of $G$ on a product $P\times F$ given by
$$ g(p,f)=(pg,g^{-1}f). $$
We denote by $N:=(P\times F)/G$ the space of orbits of this action. The bundle
$$\xi: N\to M,\quad  [(p,f)]\to \pi([p]) $$
is called an {\it associated bundle} of a principal bundle $P$. It is a vector bundle over the base manifold $M$. 
The most important examples of associated bundles of $P$ in the context of our work are the bundles with fibers $F=\mg$ and $F=G$, i.e. $N=(P\times\mg)/G$ and $N=(P\times G)/G$. From now on, we will use the notation 
$$\ad(P):=(P\times\mg)/G \quad\qquad  \textrm{and} \quad\qquad \Ad(P):=(P\times G)/G.$$ 
The action of $G$ on $\mg$ and the action of $G$ on $G$ is given by the adjoint map, namely
$$\Ad:G\times\mg\to\mg,\quad (g,X)\longmapsto\Ad_{g}(X),$$ 
$$\Ad:G\times G\to G,\quad (g,h)\longmapsto\Ad_{g}(h),   $$
respectively. For the sake of the simplicity of the notation we  have denoted both actions by the same symbol $\Ad$.


Denote by $\Omega^k(M,\mg)$ the bundle of $\mg$-valued $k$-forms on $M$. Let $\{U_\alpha\}$ be an open covering of $M$ and let $\{\xi_\alpha\}$ be a family of local $k$-forms on $M$ such that $\xi_\alpha\in\Omega^k(U_\alpha,\mg)$ for each $\alpha\in I\subset\mathbb R$. We also require that for each overlapping $U_{\alpha}\cap U_{\beta}$ the condition
\begin{equation}\label{ad}
\xi_\alpha(m)=\Ad_{g_{\alpha\beta}(m)}\circ\xi_\beta(m),   \qquad  m\in U_{\alpha\beta},\quad  g_{\alpha\beta}:U_{\alpha\beta}\to G  
\end{equation}
is satisfied. We claim that the family of $k$-forms $\{\xi_\alpha \}$ defines a $k$-form on $M$ with values in $\ad P$.  The space of $\ad P$-valued $k$-forms on $M$ will be denoted by $\Omega^k(M,\ad P)$. 

\subsection{Connection in a principal bundle}

Let $\sV P$ be a vertical subbundle in $\sT P$, i.e. a set of tangent vectors, which are tangent to the fibers of $\pi:P\to M$. A connection in a principal bundle $\pi:P\to M$ is a $G$-invariant distribution $H$ in $\sT P$, which is complementary to $\sV P$ at each point $p\in P$. By definition we have
\begin{equation}\label{dystrybhryzont}
\sT_pP=\mathsf V_pP\oplus H_p,\qquad  p\in P
\end{equation}
and
\begin{equation}\label{dystrybniezmiennicz}
H_pg=H_{pg}  \quad g\in G.
\end{equation}
The above definition is very elegant and general, however, when it comes to applications, it is more convenient to represent a connection in a different way. We will start with introducing some basic mathematical tools. Let $X$ be an element of $\mg$. The group action of $G$ on $P$ defines a vertical vector field $\sigma_X$ on $P$ associated with the element $X$, namely
$$\sigma_X(p):=\ddt p\exp(tX). $$
The field $\sigma_X$ is called a fundamental vector field corresponding to the element $X$. The fundamental vector field is equivariant in the sense that
$$\sigma_X(pg)=\sigma_{\Ad_{g^{-1}}(X)}(p).$$
A connection form in a principal bundle $P$ is a $G$-equivariant, $\mg$-valued one-form $\omega$
$$\omega:\sT P\to\mg, $$
such that $\omega(\sigma_X(p)) = X$ for each $p\in P$ and $X\in\mg$. 
The $G$-equivariance means that 
$$R^*_g\omega(p)=\Ad_{g^{-1}}\circ\omega(p) .$$
It is easy to check that the distribution $H_p:=\ker\omega(p)$ defines a connection in $P$. Since the connection form is an identity on vertical vectors, the difference of two connections is a horizontal form. It follows that the space of connections is a space of sections of an affine subbundle $\mathcal A\subset\sT^*P\otimes\mg$ modelled on a vector bundle of $\mg$-valued, $G$-equivariant horizontal one-forms on $P$. It turns out, that the space of such horizontal forms may be identified with the space of sections of the bundle $\sT^*M\otimes\ad P\to M$.

The curvature of the connection is a two-form $\Omega_\omega:=(\d\omega)^h$, where $(\d\omega)^h$ is a horizontal part of the two-form $\d\omega$. After some basic calculations one can show that
\begin{equation}\label{krzywiznaOmega}
\Omega_\omega=\d\omega+\frac{1}{2}[\omega\wedge\omega],  
\end{equation}
where $[\omega\wedge\omega]$ is a bracket of forms on $P$ with values in $\mg$. The curvature form, same as the connection form, is equivariant in a sense that $R_g^*\Omega_\omega=\Ad_{g^{-1}}\circ\Omega_\omega$. A curvature form is horizontal and $G$-equivariant therefore it defines an $\ad P$-valued two-form 

\begin{equation}\label{Fomega}
 F_\omega:M\to\wedge^2\sT^*M\otimes_M\ad P.
\end{equation} 
From now on we will use the following notation
\begin{align}
\cV & := \sT^*M\otimes_M\ad P, \label{eq:0}\\
 \cF & := \wedge^2\sT^*M\otimes_M\ad P.
\label{eq:1}
\end{align}

\subsection{Jet bundles}

Let us briefly recall the notion of first order jet spaces, which will be important in our work later on. In our presentation we will follow the notation from \cite{KG}. For a more detailed discussion of the jet bundle geometry see e.g. \cite{Saund,EG1}.

Let $\pi :E\to M$ be a smooth fibration, where $\mathrm{dim}M=m$ and $\mathrm{dim}E=m+n$. We introduce a local coordinate system $(x^i)^{m}_{i=1}$ in a domain $U\in M$. In field theory fields are represented by sections of the fibration $\pi$. The total space $E$ is the space of values of the field e.g. vector field is a section of the $\pi$ being a vector bundle, scalar field is a section of the trivial bundle $E=M\times\mathbb R$ or $E=M\times\mathbb C$, etc. On an open subset $V\subset E$ such that $\pi(V)=U$ we introduce local coordinates $(x^i,u\ind{^\alpha})$ adapted to the structure of the bundle. 

In $\sT E$ we have a vector subbundle $\sV E$ consisting of those tangent vectors that are vertical with respect to the projection $\pi$ i.e. $\sT\pi(v_p)=0$ for $v_p\in\sV_pE$. We will also need the dual vector bundle $\sV^*E$.

The space of first jets of sections of the bundle $\pi$ will be denoted by $\sJ E$.  By definition, the first jet
$\sj_m\phi$ of the section $\phi$ at the point $m\in M$ is an equivalence class of sections having the same value at the point $m$ and such that the spaces tangent to the graphs of the sections at the point $\phi(m)$ coincide. Therefore, there is a natural projection $\sj\pi$ from the space $\sJ E$ onto the manifold $E$

$$\pi_{1,0}:\sJ E\to E: \quad  \mathsf j^1_m\phi\longmapsto\phi(m).    $$
Moreover, every jet $\sj_m\phi$ may be identified with a linear map $\sT\phi:\sT_mM\to\sT_{\phi(m)}E$. Linear maps coming from jets at the point $m$ form an affine subspace in a vector space $Lin(\sT_mM,\sT_eE)$ of all linear maps from $\sT_mM$ to $\sT_eE$. A map belongs to this subspace if composed with $\sT\pi$ gives identity. In a tensorial representation we have an inclusion
$$\sJ_eE\subset\sT^*_mM\otimes\sT_eE.$$ 
It is easy to check that the affine space $\sJ_eE$ is modelled on the vector space $\sT_m^*M\otimes\sV_eE$. Summarising, the bundle $\sJ E\to E$ is an affine bundle modelled on a vector bundle 
$$\pi^*(\sT^*M)\otimes_E\sV E\to E.$$
The symbol $\pi^*(\sT^*M)$ denotes the pullback of the cotangent bundle $\sT^*M$ along to the projection $\pi$. In the following we will omit the symbol of the pullback writing simply $\sT^*M\otimes_E\sT E$ and $\sT^*M\otimes_E\sT E$.


The connection in a fibration $E\to M$ may be expressed in terms of jet bundles. Indeed, the first jet $\sj_m\phi$ defines a decomposition of the tangent space 
$$ \sT_{\phi(m)}E=\sV_{\phi(m)}E\oplus\sT\phi(\sT_mM). $$ 
Notice that $\sT\phi(\sT_mM)$ does not depend on the choice of the representative in $\sj_m\phi$. Therefore, the section of the jet bundle $\Gamma:E\to\sJ E$ defines a connection in a bundle $E\to M$. 

In the first order field theory the bundle $\sJ E$ is often called the space of infinitesimal configurations. A Lagrangian of the system is usually a map 
\begin{equation}\label{lagrang}
L:\sJ E\to\Omega^m,
\end{equation}
where $\Omega^m$ is the space of $m$-covectors on the $m$-dimensional manifold $M$. The phase space of the system is the space
$$ \mathcal P=\sV^*E\otimes_E\Omega^{m-1},  $$
which is a vector bundle over $E$. The literature concerning the mathematical formulation of the first and higher order field theory is very rich and the detailed discussion of this topic may be found e.g. in \cite{HJ1,MP2,KG,GV,JK,HJ4,LV,SG3,EG2,F7}.


\subsection{Principal connection as a section of the jet bundle}


Let us consider a first jet bundle of the principal bundle $P\to M$, i.e. a bundle $\sJ P\to P$. The action of the group $G$ on $P$ may be lifted to the action of $G$ on $\sJ P$ in a following way
$$\sJ R_g:\sJ P\to \sJ P: \quad  \sj_m\phi\to \sj_m(\phi g),  $$
where $\phi g$ is a section of $P$ such that $(\phi g)(m):=\phi(m)g$. From the above definition we obtain that the diagram
$$\xymatrix{
\sJ P   \ar[r]^{\sJ R_g}  \ar[d]_{\pi_{1,0}}  & \sJ P   \ar[d]^{\pi_{1,0}}  \\
P  \ar[r]^{R_g} &  P
} 
$$ 
is commutative. We will say that a section $\Gamma:P\to\sJ P$ is invariant under the action of $G$ if $\Gamma(pg)=\Gamma(p)g$. An invariant section of the jet bundle $\sJ P\to P$ defines a connection in a principal bundle $P$. The invariance of the section implies invariance of the horizontal distribution defined by this section. Notice, that each invariant section $\Gamma:P\to\sJ P$ defines a section
$$   \Gamma : M\to C, \qquad  \textrm{where} \qquad  C=\sJ P/G.  $$  
The bundle $C\to M$ is therefore a bundle of prinicpal connections in a principal bundle $P$. Each section of $C\to M$ defines a connection in a bundle $P\to M$. In gauge theory these sections represent gauge fields of a given theory. From the geometrical point of view $C\to M$ is an affine bundle modelled on a vector bundle $\sT^*M\otimes_M\ad P\to M$. The bundle of infinitesimal configurations for a gauge field theory is a bundle $\sJ C$, while the phase bundle is 
$$ \mathcal P=\sV^*C\otimes_C\Omega^{m-1}\simeq C\times_M\mathcal V^*\otimes_M\Omega^{m-1}\simeq  C\times_M\sT M\otimes_M\sT M\otimes_M\ad^*P\otimes_M\Omega^{m}.$$
We will introduce the notation $\barP:= C\times_M\sT M\otimes_M\sT M\otimes_M\ad^*P\otimes_M\Omega^{m}$ so that $\mathcal P=C\times_M\barP$.
It turns out that the configuration bundle $\sJ C$ has a very rich internal structure. One can show that there exists a canonical splitting over $C$
\begin{equation}\label{decompJC}
\sJ C={\mathsf J^2}P/G\oplus_C\cF,
\end{equation}
with the natural projections
$$pr_2:\sJ C\to \cF,   $$
$$(x^i, A\ind{^a_j},  A\ind{^b_j_k})\longmapsto \Big( x^i, A\ind{^a_j}, \frac{1}{2}(A\ind{^l_j_k}-A\ind{^l_k_j}+c^l_{ab}A\ind{^a_j}A\ind{^b_k}) \Big),  $$
$$pr_1:\sJ C\to {\mathsf J^2}P/G,  $$
$$(x^i, A\ind{^a_j},  A\ind{^b_j_k})\longmapsto \Big( x^i, A\ind{^a_j}, \frac{1}{2}(A\ind{^l_j_k}+A\ind{^l_k_j}-c^l_{ab}A\ind{^a_j}A\ind{^b_k}) \Big). $$
Notice, that the result of the first projection is just the curvature form of the connection
$$F_\omega=\d\omega+\frac{1}{2}[\omega,\omega],$$
which in coordinates reads
$$F_\omega=\frac{1}{2}F\ind{^a_i_j}\d x^i\wedge\d x^j\otimes e_b,   \qquad F\ind{^a_i_j}=\partial_jA\ind{^a_i}-\partial_iA\ind{^a_j}+c^a_{mn}A\ind{^m_i}A\ind{^n_j}.$$
The details concerning above decomposition may be found e.g. in \cite{SG2}. Let us notice, that the relation (\ref{decompJC}) implies that the first jet $\sj_m\omega$ of the section $\omega:M\to C$ at the point $m$ may be decomposed on the curvature of $\omega$ at the point $m$ and on some element of ${\mathsf J^2}P/G$. For a more detailed discussion of the mathematical formulation of gauge theories see e.g. \cite{EGH,LZ,SG2}.

In the end let us see that the decomposition \ref{decompJC} allows a significant simplification of the Lagrangian description of gauge theories. According to (\ref{lagrang}) the Lagrangian in gauge field theory is a map $L:\sJ C\to\Omega^m$. It is a common situation in physics that the Lagrangian of the system does not depend on the full jet of the connection but only on the value of the connection and the value of its curvature in a given point. In such a case, the Lagrangian may be reduced to a map
\begin{equation}\label{lagran2}
L:C\times\cF\to\Omega^m.
\end{equation}


\section{Gauge transformations}

In this section we will briefly recall the notion of gauge transformation, which is a cornerstone of the modern gauge field theory. Let $\pi:P\to M$ be a principal bundle with a structural group $G$. From the physical point of view, the structural group is related to the description of a gauge symmetry of the given system. The gauge transformation is an automorphism of $P$ over the identity in $M$, i.e. a diffeomorphism $P\to P$, which preserves the projection on a base manifold $M$. The set of gauge transformations forms a group with operation of composition of automorphisms as a group multiplication. We say that a system has a gauge symmetry if its Lagrangian is invariant under the action of the given group of gauge transformations. Let us briefly recall the main aspects of the theory of gauge transformations and its action on basic objects related to the principal bundle, in particular the action on connection and curvature.





\subsection{Three pictures of gauge transformations}

We will present now three equivalent pictures of the notion of gauge transformation and discuss basic properties of each of them. Let  $\pi:P\to M$ be a principal bundle with a structural group $G$. The gauge transformation of a bundle $P$ is an equivariant diffeomorphism $\Phi:P\to P$, such that the diagram
$$\xymatrix{
   P \ar[dr]_{\pi}    \ar[rr]^{\Phi}     &     & P\ar[dl]^{\pi} \\
      &     M     & 
}
$$
is commutative. The equivariance of $\Phi$ means that the condition $\Phi(pg)=\Phi(p)g$ is satisfied. Notice, that by definition each gauge transformation is an automorphism of $P$ over an identity in $M$. The set of gauge transformations (automorphisms) of the bundle $P$ will be denoted by $Aut(P)$ and by definition we have
$$ Aut(P)=\{ \Phi:P\to P, \quad \Phi(pg)=\Phi(p)g,\quad \pi\circ\Phi=\pi,  \quad p\in P,\quad g\in G  \}. $$
The set $Aut(P)$ is a group with respect to the composition of maps. It is easy to check that if $\Phi_1$ and $\Phi_2$ are automorphisms of $P$ then the composition $\Phi_1\circ\Phi_2$ is an automorphism of $P$ as well. Each automorphism of a principal bundle over the identity on $M$ may be written in the form
$$\Phi(p)=pf(p), \quad f:P\to G,  $$
where $f$ is a unique function on $P$ with values in $G$ associated with $\Phi$. The equivariance of the map $\Phi$ implies the condition
\begin{equation}\label{funkcjaf}
f(pg)=\Ad_{g^{-1}}f(p).
\end{equation}
It is easy to check that if $\Phi_1$, $\Phi_2$ are realised by the functions $f_1, f_2$ respectively, then
$$\Phi_1\circ\Phi_2(p)=pf_1(p)f_2(p),  $$
which means that the gauge transformation $\Phi_1\circ\Phi_2$ is realised by the function $f_1\cdot f_2$. The set of gauge transformations understood as functions on $P$ with values in $G$ form therefore a group with respect to the multiplication of functions. We denote this group by
\begin{equation}\label{grupacechow}
\mathcal G:=\{ f:P\to G, \quad f(pg)=\Ad_{g^{-1}}f(p),   \quad p\in P,\quad g\in G    \}.
\end{equation}
On the other hand, each function on $P$ with values in $G$ is a section of the trivial bundle $P\times G\to P$ given by the formula
$$ \sigma_f:P\to P\times G, \quad p\longmapsto (p, f(p)).$$
From the fact that the function $f$ satisfies the condition (\ref{funkcjaf}) we obtain that the section $\sigma_f$ defines a section of $\Ad P\to M$ given by the formula
\begin{equation}\label{cieciesigmaf}
\bar\sigma_f:M\to\Ad P, \quad m\longmapsto [p, f(p)], \quad \pi(p)=m, 
\end{equation}
where $[p, f(p)]$ is an equivalence class of the element $(p, f(p))$ in $\Ad P$. The set $\Gamma(\Ad P, M)$ of sections of the bundle $\Ad P\to M$ forms a group with respect to the multiplication of sections.

From above considerations we obtain that gauge transformation may be seen in three different ways: (1) as an automorphism of a principal bundle, (2) as a function on $P$ with values in $G$ satisfying (\ref{funkcjaf}) and (3) as a section of the bundle $\Ad P\to M$. In our subsequent work we will mostly rely on the approach (1) and (2).  


\subsection{Gauge transformation of the connection and curvature}

The group of gauge transformations acts in a natural way on connection, curvature and covariant derivative in a bundle $P$. Let $H$ be a principal connection in a bundle $P$, i.e. a distribution in $\sT P$ satisfying the conditions (\ref{dystrybhryzont}) and (\ref{dystrybniezmiennicz}). The gauge transformation of a connection $H$ is a distribution
$$ H^\Phi:=\Phi_*H.$$
It is easy to check that $H^\Phi$ satisfies (\ref{dystrybhryzont}) and (\ref{dystrybniezmiennicz}), therefore it defines a connection in $P$. In terms of a connection form, the distribution $H^\Phi$ is related to the connection form $\omega^\Phi:=\Phi^*\omega$. Let us take a closer look on a form $\omega^\Phi$. Let $v\in\sT_pP$ be a tangent vector represented by a curve $\gamma$, and $\Phi(p)=pf(p)$ a gauge transformation. It is a matter of computation to check that
\begin{equation}\label{transformomega}
\Phi^*\omega(p)=\Ad_{f(p)^{-1}}\circ\omega+f(p)^{-1}\d f(p). 
\end{equation}
Notice, that in the above formula $f$ is a function with values in $G$, which means that $\d f(p)\in\sT_p^*P\otimes\sT_{f(p)}G$ and $f(p)^{-1}\d f(p)\in\sT_p^*P\otimes\mg$. 

The action of gauge tranformation may be similarly extended to other objects associated with the principal bundle. Let $\Omega$ be a curvature of the connection $\omega$ defined as in (\ref{krzywiznaOmega}), and $D_\omega\alpha$ a covariant derivative of a section $\alpha:M\to\wedge^k\sT^*M\otimes_M\ad P$ with respect to a form $\omega$ and representation $\rho$. Performing a similar calculation as for $\Phi^*\omega$ one can derive the transformation formulas
$$\Phi^*\Omega=\Ad_{f^{-1}}\circ\Omega,   $$
$$\Phi^*D_\omega\alpha=\rho(f^{-1})D_\omega\alpha.   $$

\section{Reduction of gauge symmetries - the dressing field method}

Let us briefly present the main features of the dressing field method approach to the reduction of a gauge symmetry. A detailed discussion of the notion of the dressing field together with numerous examples may be found in \cite{AT,AFM,FFLM,FLM,MW}. The existence of the gauge symmetry means that the Lagrangian is invariant under the action of the group of gauge transformations describing the symmetry. The main idea behind the dressing field method is to introduce a certain auxiliary field, which, in a general case, does not belong to the original space of gauge fields. We call this auxiliary field a {\it dressing field}. In the next step, we define a transformation of the gauge fields that depends on the dressing field on the one hand and on the original gauge fields on the other. It turns out that in certain situations these new fields are invariant under the action of the gauge group (or its subgroup), which in turns means that the symmetry of the system has been (fully or partially) reduced.

We will move now to the technical aspects of the dressing field method. Let $\pi:P\to M$ be a principal bundle with a structural group $G$. Assume that we have a distinguished subgroup $H$ in $G$. We recall the gauge group of the theory (\ref{grupacechow}), given by 
$$\mathcal G:= \Big\{  f: P\to G, \quad  f(pg)=\Ad_{g^{-1}}f(p), \quad  g\in G  \Big\}.  $$
We introduce now a following set of maps associated with the subgroup $H$
$$\mathcal H:= \Big\{  f^\mathcal H: P\to H, \quad  f(ph)=\Ad_{h^{-1}}f(p), \quad h\in H\subset G \Big\}.  $$
Notice that each element of $\mathcal H$ defines a map
\begin{equation}\label{PhiH}
 \Phi^{\mathcal H}:P\to P, \quad \Phi^{\mathcal H}(p)=pf^\mathcal H(p), 
\end{equation}
which, despite the similarity of the notation, is not a gauge transformation. Let us introduce a function
$$u:P\to H,  $$  
which transforms with respect to $H$ according to the formula
\begin{equation}\label{utransform}
u(ph)=h^{-1}u(p).  
\end{equation}
The function $u$ will be called a {\it dressing field}. It is a crucial object in the whole dressing field method. Let us stress that the condition (\ref{utransform}) implies that $u$ is not an element of $\mathcal H$. In the next step we use $u$ to define a map 
\begin{equation}\label{PhiU}
\Phi^u:P\to P,\quad \Phi^u(p)=pu(p).
\end{equation}
It is easy to check that it satisfies the relation $\Phi^u(ph)=\Phi^u(p)$. Notice, that the above map is not bijective, therefore it does not define a gauge transformation. However, the existence of $\Phi^u$ provides a decomposition of the bundle $P$. Notice that the condition $\Phi^u(ph)=\Phi^u(p)$ implies that $\Phi^u$ is constant on the orbits of action of the subgroup $H$. It means that $\Phi^u$ defines a section of the bundle $P\to P/H$. The global section of the principal bundle uniquely defines a global trivialisation of that bundle. Therefore, we obtain that $P=P/H\times H$. It turns out, that there exists the opposite implication as well, i.e. the decomposition $P=P/H\times H$ defines a suitable dressing field. Indeed, one can show that the existence of the field $u:P\to H$ is equivalent to the existence of the decomposition $P=P/H\times H$ \cite{F}.    

The map $\Phi^u$ naturally acts on the space of principal connections. Let $\omega$ be a connection form on $P$. We say that the map $\omega^u$ given by
$$ \omega^u:\sT P\to\mg,  \qquad   \omega^u:={\Phi^u}^*\omega.$$ 
is a dressing of the connection form $\omega$. It is a matter of straightforward computations to derive the formula
\begin{equation}\label{transfkonek}
\omega^u=\Ad_{u^{-1}}\circ\omega+u^{-1}\d u. 
\end{equation}
Using (\ref{transfkonek}) one can easily show that the form $\omega^u$ is invariant under the action of $H$. Furthermore, it is also invariant under the transformations given by (\ref{PhiH}). Indeed, for each $f\in\mathcal H$ by definition we have
$$ \Phi^{\mathcal H*}\omega^u=(\Phi^u\circ\Phi^{\mathcal H})^*\omega. $$
On the other hand 
$$  \Phi^u\circ\Phi^{\mathcal H}(p)=\Phi^u(pf^\mathcal H(p))= pf^\mathcal H(p)u(pf^\mathcal H(p))=  pf^\mathcal H(p)f^\mathcal H(p)^{-1}u(p)=pu(p)=\Phi^u(p), $$
which implies that 
$$\Phi^{\mathcal H*}\omega^u={\Phi^u}^*\omega=\omega^u.$$
Notice that in the above calculations we have used (\ref{utransform}), which means that it is valid only for functions $f$ with values in $H$.

Let us emphasize two particular features of the above construction. First of all, the dressing transformation given by (\ref{PhiU}) is not a gauge transformation, despite the algebraic similarity of both maps. It means that the fields $\omega^u$ and $\omega$, in a general case, do not belong to the same orbit of the action of $\mathcal G$. Therefore, the fields $\omega^u$ are not, in general, elements of the original space of gauge fields. Secondly, the form of the dressing field has to be deduced ad hoc, basing on the specific form of the Lagrangian.

In the end, let us see the above constructions in application to the example coming from the theory of electroweak interaction. 
Let $\rho$ be a representation of the group $G=SU(2)\times U(1)$ on $\C2$ given by
$$ \rho: G\to End(\C2), \quad \rho(b,a)v=bav.  $$
We denote the restrictions of $\rho$ to $U(1)$ and $SU(2)$ by $\rho_1$ and $\rho_2$, respectively. Let $E$ be an associated bundle of $P$ with respect to the above representation and with a typical fiber $\C2$, i.e. $E:=P\times_\rho\C2$. 
We consider a section $\phi:M\to E$, which, by definition, defines a section of the trivial bundle $\bar\phi:P\to P\times\C2$ satisfying the condition
\begin{equation}\label{warunekphi}
\qquad\qquad \bar\phi(pg)=g^{-1}\phi(p), \qquad g=(b,a)\in SU(2)\times U(1), 
\end{equation}
where $g^{-1}\phi(p)$ is a matrix multiplication of a vector $\bar\phi(p)\in\C2$ by a pair of matrices $g^{-1}\in SU(2)\times U(1)$. The group $SU(2)$ is a subgroup in $SU(2)\times U(1)$. Section $\bar\phi$ defines a map $u:P\to SU(2)$ given by the formula
\begin{equation}\label{dressingprzyklad}
\bar\phi(p)=u(p)\eta, \qquad \textrm{where} \qquad  u:P\to SU(2), \quad  \eta=\begin{pmatrix} 0 \\ ||\bar\phi ||  \end{pmatrix}.
\end{equation}
It is easy to check that the condition (\ref{warunekphi}) implies that $u(pb)=b^{-1}u(p)$ for $b\in SU(2)$, which means that $u$ is indeed a dressing field.

\section{Geometric approach to the dressing field method}

In this section we will present the main result of our paper, which is a geometric description of the dressing field method in a presence of the residual symmetry of $u$. We will consider the situation when $u$ transforms under $J$ with respect to the adjoint action and when there is a decomposition $G=JH$. In particular we will focus on a case when $G$ is a direct product of $H$ and $J$. It is one of the two main situations originally considered in \cite{AFM}. Such a situation occurs for instance in the electroweak theory where the structural group is $SU(2)\times U(1)$. We will show how the geometry of the underlying principal bundle is affected by the existence of such a dressing field and, as a consequence, how the configuration and phase bundle of the theory may be reduced.

\subsection{Dressing map as a principal bundle}

Let us assume that there exists distinguished subgroups $J$ and $H$ in $G$, such that $H$ is a normal subgroup in $G$ and each element of $G$ may be uniquely written in a form $g=jh$, i.e. $G=JH$. Such a situation occurs for instance when $G$ is a direct or a semidirect product of $H$ and $J$. Then $G/H$ has a structure of a Lie group as well and $G/H\simeq J$. The Lie algebra $\mj$ of the group $J$ is isomorphic to $\mg/\mh$. The bundle
$$G\to J$$
is a principal bundle with a structural group $H$. Let us introduce the notation $P^J:=P/H$. The fibrations
$$\pi_{P^J}:P\to P^J, $$ 
$$\pi_J:P^J\to M,$$ 
are principal bundles with structural groups $H$ and $J$, respectively. 

Let us fix now a chosen dressing field $u$ on $P$. It defines an embedding of $P^J$ in $P$ given by
$$ P\supset P^J=u^{-1}(e).$$
Notice, that $u(\Phi^u(p))=e$, so the preimage $u^{-1}(e)$ is an image of the section $P\to P/H$ defined by $\Phi^u$. The set $P^J$ is therefore a submanifold in $P$, on which $\Phi^u$ acts in a trivial way. The above embedding defines a trivialisation of $P$ given by
\begin{equation}\label{rozkladPJ}
P\to P^J\times H, \quad  p\longmapsto (p_0, h), \quad \textrm{where} \qquad p_0=ph, \quad h=u(p).
\end{equation}
Let us notice that $u(p_0)=u(ph)=h^{-1}u(p)=e$, so that $ph\in P^J$. One can check how $\Phi^u$ behaves under above decomposition. Let $p\in P$ be a point, which in the identification (\ref{rozkladPJ}) has a form $(p_0,h)$. Acting $\Phi^u$ on $p$ we get
$$\Phi^u(p)= \Phi^u(p_0h^{-1})= \Phi^u(p_0)=p_0. $$ 
From the above calculation we obtain that $\Phi^u$ is a projection on a submanifold $P^J$, i.e. $\Phi^u=\pi_{P^J}$. Therefore, it turns out that in a case when a structural group has a decomposition $G=JH$, the existence of the dressing field $u$ is equivalent to the existence of the embedding $P^J\simeq P/H$ in $P$, and, as a consequence, to the existence of the decomposition $P=P^J\times H$ over $M$.

\subsection{Residual gauge symmetry and reduced connection form}

The gauge symmetry of the dressed fields depends on the original gauge fields on the one hand and on the dressing field on the other. In practical applications it is usually important how $u$ transforms with respect to the action of $J$. In the following part of this section we will analyse the case
\begin{equation}\label{Jtrsfm}
R^*_ju=\Ad_{j^{-1}}u,
\end{equation}
which has applications in a BRST differential algebra. The algebraic discussion of the above case may be found in \cite{AFM}. Our aim is to analyse how (\ref{Jtrsfm}) affects the geometry of the principal bundle, on which $u$ is defined. 

\noindent Let us notice, that the condition (\ref{Jtrsfm}) implies
$$ \Phi^u(pj)=pju(pj)=pu(p)j=\Phi^u(p)j, \qquad p\in P, \quad j\in J,   $$
which means that $\Phi^u$ commutes with the right action $R_j$. We recall, that a dressed connection is a one-form $\omega^u={\Phi^u}^*\omega$. The connection form $\omega$ restricted to tangent vectors $\sT P^J\subset\sT P$ defines a connection form in a principal bundle $P^J\to M$. Let us denote the restriction of $\omega$ to $\sT P^J$ by $\omega^J$. From the conclusions of the previous section, i.e. $\Phi^u=\pi_{P^J}$, and the fact that $P^J$ is a submanifold in $P$ we obtain, that the dressing of the connection form is just a pull-back of this connection form with respect to the projection $\pi_{P^J}$. In the obvious way we have $\pi_{P^J}^*\omega=\pi_{P^J}^*\omega^J$, so that we obtain $\omega^u=\pi_{P^J}^*\omega^J$. Notice, that $\omega^u$ is also equivariant with respect to the action of $J$, which comes from the fact that
$$ R_j^*\omega^u=R_j^*\pi_{P^J}^*\omega^J=\pi_{P^J}^*R_j^*\omega^J=\pi_{P^J}^*\Ad_{j^{-1}}\circ\omega^J= \Ad_{j^{-1}}\circ\omega^u. $$

From the above discussion we conclude, that if $u$ satisfies the condition (\ref{Jtrsfm}), then the form $\omega^u$ is a pull-back of the connection one-form in a principal bundle $\pi_J:P^J\to M$. 





\subsection{ Adjoint bundle of the reduced principal bundle}

In the following we will assume that $P$ is equipped with a fixed dressing field $u$, i.e. there exists a decomposition $P=P^J\times H$. For the sake of the clarity of the presentation we will introduce the notation $\wdt P:=P^J$, $\wdt\pi:=\pi^J$. The Lie algebras of the groups $J$ and $H$ will be denoted by $\mathfrak j$ and $\mathfrak h$ respectively. Since the existence  of the field $u$ is equivalent to the existence of the embedding $\wdt P\hookrightarrow P$, from now on we will understand the dressing field method rather as a choice of the suitable embdedding of $\wdt P$ in $P$ than as a map $u:P\to H$.

Let us consider the adjoint bundle of the principal bundle $\wdt \pi:\wdt P\to M$. The structural group of $\wdt P$ is the subgroup $J$, so by definition we have
$$\ad\wdt P:= (\wdt P\times\mj )/J.  $$
The decomposition $G=H\times J$ implies the decomposition $\mg=\mh\oplus\mj$. We will denote the projection from $\mg$ onto $\mj$ by $pr_\mj$. There exists a canonical projection
$$ \pi_{\wdt P}\times pr_{\mj}: P\times\mg\to\wdt P\times\mj, $$
$$ (p,X)\to \Big(\pi_{\wdt P}(p), pr_\mj(X) \Big).  $$
If we divide both sides of the above projection by $G$ and use the fact that $H$ acts on $\wdt P$ in a trivial way we will obtain a projection
\begin{equation}\label{Delta1}
\Delta:\ad P\to \ad\wdt P, \qquad [p,X]_G\longmapsto [\pi_{\wdt P}(p), pr_\mj(X)]_J, 
\end{equation}
where $[p,X]_G$ is an equivalence class of the element $(p,X)$ with respect to the action of $G$, and $[\pi_{\wdt P}(p), pr_\mj(X)]_J$ is an equivalence class of the element $(\pi_{\wdt P}(p), pr_\mj(X))$ with respect to the action of $J$. One can easily check that the above projection does not depend on the choice of the representative. Indeed, we have
$$\pi_{P^J}\times pr_{\mj}(pg,\Ad_{g^{-1}}\circ X) = \Big(pr_{P^J}(pg), pr_\mj(\Ad_{g^{-1}}\circ X) \Big)= \Big(pr_{P^J}(p)j, \Ad_{j^{-1}}\circ pr_\mj( X) \Big) $$
where $j$ is a projection of $g$ onto $J$. Notice that the equality $pr_\mj(\Ad_{g^{-1}}\circ X)=\Ad_{j^{-1}}\circ pr_\mj(X)$ requires the existence of a direct product structure in $G$. Therefore, the map (\ref{Delta1}) is not well-defined in a more general case, e.g. when $G$ is a semi-direct product of $H$ and $J$. Using above formula we obtain a projection
\begin{equation}\label{Delta2}
\Delta:\Omega^k(M)\otimes_M\ad P\to\Omega^k(M)\otimes_M\ad\wdt P, 
\end{equation} 
which for the clarity of the notation we have denoted by the same symbol as (\ref{Delta1}). Above map may be lifted to a map $\sJ\Delta$ acting on jet bundles and represented by a diagram
\begin{equation}\label{JDelta} \hspace*{+1cm} 
\xymatrix@C+30pt{
\sJ\Big(\Omega^k(M)\otimes_M\ad P\Big) \ar[d]^{}  \ar[r]^{\sJ\Delta}    &  \sJ\Big(\Omega^k(M)\otimes_M\ad\wdt P\Big)  \ar[d]^{}    \\
  \Omega^k(M)\otimes_M\ad P   \ar[r]^{\Delta}  \ar[d]^{pr_M}  & \Omega^k(M)\otimes_M\ad\wdt P \ar[d]^{pr_M} \\
M   \ar[r]^{id}  &M
} 
\end{equation}


\subsection{Reduction of the connection bundle}

Let us consider a bundle of first jets of sections of $\wdt\pi$, i.e. a bundle $\sJ\wdt P\to\wdt P$. The action of $J$ on $\wdt P$ may be lifted to the action on the total space $\sJ\wdt P$. Principal connections in $\wdt P$ are represented by sections of the bundle
\begin{equation}\label{tildeC}
\wdt C\to M,
\end{equation}
where $\wdt C:=\sJ\wdt P/J$. The bundle (\ref{tildeC}) is an affine bundle modeled on a vector bundle $\sT^*M\otimes_M\ad\wdt P\to M$. The isomorphism 
$$ P\to \wdt P\times_MH    $$
may be lifted to the isomorphism
$$ \sJ P\to \sJ\wdt P\times_M\sJ(M\times H)    $$
over $M$. In particular, the projection $\pi_{\wdt P}:P\to\wdt P$ defines a map $\sJ\pi_{\wdt P}:\sJ P\to\sJ\wdt P$ between jet bundles over $M$ and represented by a diagram
\begin{equation}\label{diagramJP}
\xymatrix@C+20pt{
{\sJ P} \ar[d]^{pr_P}  \ar[r]^{\sJ\pi_{P^J}}    &   \sJ\wdt P   \ar[d]^{pr_P}    \\
 P    \ar[d]^{pr_M}   \ar[r]^{\pi_{P^J}}   & \wdt P  \ar[d]^{pr_M}   \\
        M    \ar[r]^{id}   &    M
} 
\end{equation}
Let us consider now a section $\omega:P\to\sJ P$. Notice that the map
$$\wdt\omega:\wdt P\to\sJ\wdt P, \qquad \wdt\omega(\wdt\pi(p))=\sJ\wdt\pi(\omega(p)) $$
is in general not well-defined. It is easy to check, that for $p_2=p_1h$, where $h\in H$, we have $\wdt\pi(p_2)=\wdt\pi(p_1)$, but $\sJ\wdt\pi(\omega(p_2))\neq \sJ\wdt\pi(\omega(p_1))$. Let us impose the additional condition that $\omega$ is $G$-invariant, i.e. $\omega(pg)=\omega(p)g$. Then for $p_2=p_1g$, where $g=hj$, we have $\wdt\pi(p_2)=\wdt\pi(p_1g)=\wdt\pi(p_1)j$. Finally, we obtain
$$ \wdt\omega(\wdt\pi(p_2))=\sJ\wdt\pi(\omega(p_2))=\sJ\wdt\pi(\omega(p_1g))= \sJ\wdt\pi(\omega(p_1)g)=\sJ\wdt\pi(\omega(p_1))j,   $$
and as a consequence
$$ \wdt\omega(\wdt\pi(p_1)j)=\wdt\omega(\wdt\pi(p_1))j.  $$
It turns out that for $\omega$ being a $G$-invariant section the map $\wdt\omega$ is a well-defined $J$-invariant section. The map $\wdt\pi$  defines therefore a projection of the $G$-invariant section of the bundle $\sJ P\to P$ onto a $J$-invariant section of the bundle $\sJ\wdt P\to\wdt P$. Similarly, one can show that each $G$-invariant section $\omega$ defines a $H$-invariant section of the bundle $M\times H\to\sJ(M\times H)$. In particular, by dividing the left-hand side of the diagram (\ref{diagramJP}) by $G$ and the right-hand side by $J$ we obtain a diagram
\begin{equation}\label{diagramC}
\xymatrix@C+20pt{
C   \ar[d]^{pr_M}  \ar[r]^{\delta}    &  \wdt C   \ar[d]^{pr_M}    \\
        M    \ar[r]^{id}   &    M
} 
\end{equation}
The map $\delta$ defines a reduction of the connection bundle in a given gauge theory. The first jet prolongation of $\delta$
\begin{equation}\label{dzetdelta}
\sJ\delta:\sJ C\to \sJ\wdt C.
\end{equation} 
provides a reduction of the configuration bundle.
Notice that in the light of the discussion above the map $\delta$ defines also a projection
$$\Gamma(C,M)\to \Gamma(\wdt C, M) $$ 
between modules of sections of bundles $C\to M$ and $\wdt C\to M$.
If $\omega$ is a section of the bundle $C\to M$, then the map
$$ \wdt\omega:M\to\wdt C, \quad \wdt\omega:=\delta\circ\omega $$
is a section of the bundle $\wdt C\to M$. 
Let us emphasize that $\delta$ depends on the choice of a dressing field $u$, which comes from the fact that $\Phi^u=\pi_{\wdt P}$.


\subsection{ Reduction of the configuration and phase bundle}

The reduction of the connection bundle implies a reduction of the configuration and phase bundle. Let us introduce the notation
$$\wdt \cV:=\sT^*M\otimes_M\ad\wdt P,  $$
$$\wdt \cF:=\wedge^2\sT^*M\otimes_M\ad\wdt P.$$
Using (\ref{decompJC}) and (\ref{dzetdelta}) we obtain decompositions
$$\sJ C=\J2 P/G\oplus_C(C\times_M\cF),    $$
$$\sJ\wdt C=\J2\wdt P/J\oplus_{\wdt C}(\wdt C\times_M\wdt \cF).    $$
represented by a diagram
$$\xymatrix{
\sJ C \ar[d]^{\sJ\delta}  \ar[r]    &  \J2 P/G\oplus_C(C\times_M\cF)  \ar[d]^{pr}    \\
 \sJ\wdt C  \ar[r]   & \J2\wdt P/J\oplus_{\wdt C}(\wdt C\times_M\wdt \cF) \\
} 
$$
Above decompositions and the projection $\sJ C\to\sJ\wdt C$ defines a map
$$ \J2 P/G\to \J2\wdt P/J, $$
and
\begin{equation}\label{CtimesF}
\zeta:C\times_M\cF\to \wdt C\times_M\wdt \cF, \quad (\omega, F)\longmapsto (\wdt\omega,\wdt F) .
\end{equation}
In particular, we will be interested in the map (\ref{CtimesF}). Let us recall that the phase bundle of a gauge theory is a bundle $C\times\overline{\mathcal P}\to C$, where $\barP:= C\times_M\sT M\otimes_M\sT M\otimes_M\ad^*P\otimes_M\Omega^{m}$. Applying the map (\ref{Delta1}) to $\barP$ we obtain
$$\Delta:\barP\to\wdt{\mathcal P}, \quad\bar p\longmapsto\wdt p, $$
where 
$$ \wdt{\mathcal P}:= C\times_M\sT M\otimes_M\sT M\otimes_M\ad^*{\wdt P}\otimes_M\Omega^{m}$$
and $\ad^*\wdt P\to M$ is a dual bundle to the bundle $\ad\wdt P\to M$. The reduced phase bundle is therefore a bundle $\wdt C\times\wdt{\mathcal P}\to\wdt C$. Similarly, for the bundle $\sJ\overline{\mathcal P}$ we obtain the reduction
$$ \sJ\Delta:\sJ\barP\to\sJ\wdt{\mathcal P}, \quad \sj_m\bar p\longmapsto\sj_m\wdt p.  $$
In the above framework one can also include a reduced Lagrangian. From (\ref{lagran2}) we have obtained that the Lagrangian of a gauge theory has a form
$$ L:C\times_M \cF\to\Omega^m.   $$
Let us assume now that $L$ depends only on the projection on $\wdt C\times_M\wdt \cF$. Then, we can introduce a reduced Lagrangian
$$ \tilde L:\wdt C\times_M\wdt\cF\to \Omega^m, \quad   L =\tilde L\circ\zeta.   $$
From the above considerations we conclude that by introducing dressing fields adapted to the form of the Lagrangian of the system, we can partially, or fully, reduce the gauge symmetry of the given theory. Let us notice that if the Lagrangian has a gauge symmetry described by the group $G$, then the solutions of the equations of motion, by definition will have this symmetry as well. It means that if the field $\omega$ is a solution of the equations of motion then each field $\omega^\Phi$ is also such a solution. By performing a reduction of the symmetry with respect to the subgroup $H\subset G$ and by introducing a reduced Lagrangian $\tilde L$ we obtain new equations of motion, which are invariant under the gauge transformations described by the subgroup $J$. The corresponding solutions of the reduced equations of motion will have a symmetry given by $J$ as well.


\section{Acknowledgments}

Th author gratefully acknowledge dr. hab. Katarzyna Grabowska for her valuable comments, which definitely make this paper more correct and easy to read.

\end{document}